\documentstyle[prl,aps,psfig]{revtex}

\begin{document}


\draft
\preprint{Universit\'e Laval preprint LAVAL-PHY-07/95}

\title{Computation of Structure Functions
 From a Lattice Hamiltonian}

\author{H. Kr\"oger and N. Scheu}

\address{
D\'epartement de Physique, Universit\'e Laval,
Qu\'ebec, Qu\'ebec G1K 7P4,
Canada.
E-mail: nscheu@phy.ulaval.ca
}

\date{\today}

\maketitle

 \begin{abstract}

We compute structure functions in the Hamiltonian formalism
on a momentum lattice using a physically motivated regularisation that links
the maximal parton number to the lattice size.
We show for the $\phi ^4 _{3+1}$ theory that our method allows to describe
continuum physics. The critical line and the renormalised mass spectrum close
to the critical line are computed and scaling behaviour is observed
in good agreement with L{\"u}scher and Weisz' lattice results.
We then compute distribution functions and find a $Q^2$ behaviour
and the typical peak at $x_B\rightarrow 0$ like in $QCD$.

\end{abstract}



Hadron structure is probed by deep inelastic scattering ($DIS$).
Over recent years a great deal of experimental data has been
gathered from high energy collider experiments.
While perturbative quantum
chromodynamics ($QCD$) describes successfully the large $Q^2$ dependence
of $DIS$ structure functions, it fails to predict the correct dependence
on the Bj{\o}rken variable $x_B$.
Thus much
effort has been devoted to
compute quark or gluon distribution
functions and proton structure functions from $QCD$ with
{\it non-perturbative} methods. E.g., Martinelli
et al.\cite{Martinelli} have computed the first two moments of the
pion structure function via Monte Carlo lattice simulations.
These calculations are notoriously difficult (for
the present status of lattice calculations of structure functions
see Ref.\cite{Lattice}).
This situation calls for alternative techniques.
In this letter we present such a new approach. Its basic ingredients
are: (i) We use a Hamiltonian formulation, based on
(ii) a momentum lattice as regulator, and
(iii) use a Breit frame ({\it not} the rest frame) corresponding to the
scattering process.
We apply our method to the scalar model in $3+1$ dimensions, which
has been extensively studied,
and compute the distribution function. As a result we find an Altarelli-Parisi
like behaviour leading to a sharp forward peak at small $x_B$
at high resolution $Q^2$, as it typically shows up in high energy $DIS$ hadron
scattering experiments.
We extract continuum physics:
Close to the critical point our results are in perfect agreement with the
predicted scaling behaviour as well as with Euclidian lattice  results
by L\"uscher and Weisz\cite{Luscher}.

\bigskip

Let us briefly outline the reasons for the choice of our method:
(i) Structure functions are computed from wave functions. Wave functions
are defined in Minkowsky space where they can be computed directly from
a Hamiltonian formulation.
The Hamilonian approach offers the advantage of allowing to
compute directly Minkowsky space
observables. E.g., scattering wave
functions for glueball-like states in compact $QED_{2+1}$ have been computed
in a Hamiltonian formulation on a momentum
lattice\cite{ChaaraHelmut} (for a review of Hamiltonian lattice methods
see \cite{HelmutsReview,Schutte,BrodskyPauli}).
(ii) The usefulness of a momentum lattice
to compute physics close to a critical point  has
been demonstrated in Ref. \cite{HelmutStephane,Espriu}.
(iii) The reason for our choice of the Breit frame will be explained below.
However, Hamiltonian methods are known to lead to numerical
problems because of the
huge number of degrees of freedom involved.
Nobody has succeeded yet in observing scaling
behaviour indicating continuum physics in a (3+1)-dimensional Hamiltonian
formulation.
In this work we shall demonstrate for the scalar theory
that those difficulties can be
overcome.

\bigskip

The most important experiment in order to probe the structure of hadrons
is deep inelastic scattering ($DIS$)). Its simplest form is inclusive
scattering of an
unpolarised lepton off a hadronic target. Let us recall some basic
notations \cite{Roberts}.
The hadron in its ground state with four momentum $P$ interacts with
the probing lepton by the exchange of a virtual photon (our neutrino)
with space-like four-momentum $q$.
In Feynman's parton
model it is assumed that the proton consists of constituents,
the partons, which are weakly bound, i.e. its binding energy is small
compared to the resolution ability
$Q:=\sqrt{-q_\mu q^\mu}$ of the probing photon.
In this approximation, the so-called Bj{\o}rken scaling variable
 $x_B:= {Q^2\over 2P_\mu q^\mu}$ can be
interpreted as the momentum fraction of the struck parton if we work in the
Breit frame. The Breit frame is defined by the requirements that the photon
energy $q_0$ be zero and
that the photon momentum $\vec q$ be antiparallel to the hadron
momentum $\vec P$. In this frame the following relation
between the parton momentum $\vec p$ and the proton momentum holds:
\begin{equation} \label{criterion}
(\vec p-\vec P/2)^2 \leq |\vec P/2|^2   .
\end{equation}
The rationale for this particular choice of frame being that
$QCD$ structure functions $F(x_B,Q)$ can be
interpreted as a linear combination of parton momentum distribution
functions $f(x_B,Q)$, which have a more intuitive interpretation. The latter
is defined by th Structure
functions are another way of expressing scattering cross sections.
The distribution function of a parton counts the number of those partons
with a given momentum fraction $x_B$ in the proton. For a precise definition
see Ref.\cite{Roberts}.

\bigskip

Because the Breit frame introduced above refers to a particular struck
parton and we want to describe a many-parton system (proton) we need
to extend the definition to a generalised Breit frame: $q_0=0$ but
$\vec q$ needs no longer be antiparallel to $\vec P$. Although the parton
momentum needs no longer be collinear in general to the proton
momentum $\vec P$, we nevertheless
impose Eq.(1) as kinematical condition.
While the generalised Breit frame has been introduced for the purpose of
practical calculations, it should be noted that the
strict relation between distribution and structure functions,
characteristic for original Breit-frame no longer holds in a strict
sense. However, this relation is recovered for the generalised
Breit frame in the continuum limit.

\bigskip

Because we are working in the Hamiltonian approach we need to define
a basis of the Hilbert space.
We construct the Hilbert space as a Fock space of free particles
and select (parton) momenta $\vec p$
from a bounded domain corresponding to
$DIS$ as given by Eq.(1).
This is an {\it assumption} based on the physical intuition that
the experimentally observable
parton momenta are those which dominate the quantum
dynamics. This assumption has been tested by computing critical
behaviour of renormalised masses and a good agreement with
analytical scaling behaviour has been observed (see below).

\bigskip

Now we introduce a momentum lattice regularisation:
In order to have a practically convenient lattice we further
constrain the parton momenta from Eq.(1), namely by selecting a regular cube
centered at $\vec P/2$ and located inside the ball given by Eq.(1).
I.e., the parton momenta $\vec p$ lie in the domain
\begin{equation}
0\leq p_i \leq \Lambda= {\sqrt{3} \over 2} |\vec P|
    \;\;\; \mbox{for}\; i=x,y,z.
\end{equation}
We define lattice momenta
$\vec p:= \vec n \Delta p$ where $\vec n$
is an integer vector
and ${\Delta p}$ is the momentum lattice spacing covering the domain
given by Eq.(2).
One notices that all lattice momenta are $positive$ (non negative).
Contrary to a regularisation in the rest frame which does {\it not}
limit the particle number, our approach has the following important
property: For any given Hilbert state with non-zero total momentum,
the Fock space particle numbers are bounded. Consequently the ultraviolet
cutoff $\Lambda$ given by Eq.(2) implies a total particle number cutoff
and thus drastically reduces the dimension of the Hilbert space.
\\
\\
{\Large\bf Mass spectrum and critical behaviour of the $\phi^4 _{3+1}$ theory}
 \\
Before discussing structure functions we need to convince ourselves
that the method allows to compute correctly physical observables. We have
chosen the scalar $\phi^4 _{3+1}$ theory because it is a quite well
understood theory and has a second order phase transition,
allowing to test our method near a critical point.
The Hamiltonian of the $\phi^4$ theory
is given by
\begin{equation}
H= \int d^3 x \; \frac{1}{2} ({\partial \phi \over \partial t})^2 +
  \frac{1}{2}(\vec \nabla \phi)^2 + {m_0^2\over 2} \phi^2 + {g_0\over 4!}
\phi^4,
\end{equation}
where $m_0$ and $g_0$ are the bare mass and
coupling constant, respectively.
We express the Hamiltonian in terms of free field creation and annihilation
operators corresponding to the lattice momenta.
Because the Hamiltonian and the momentum operators commute, we compute
the energy spectrum $E_n$ in a Hilbert space sector
of given momentum $\vec P$.
Since we are not in the rest frame
we have to use the mass-shell condition
$M_n:=\sqrt{E_n ^2-\vec P ^2}$ in order to obtain the physical mass spectrum.
It is known\cite{Luscher}
that the critical line between the symmetric and the broken phase
lies entirely in the region where the bare parton
mass squared $m_0^2$ is negative.
Hence we cannot build up the Fock-space in terms of
partons with those masses.
As a remedy we have split the bare mass squared
$m_0^2 = m_{kin} ^2 + m_{int} ^2$
into a positive kinetic part $m_{kin}^2$ and an interaction part $m_{int} ^2$.
The Fock states are built from positive bare masses $m_{kin}$.
The best choice of $m_{kin}$ seems to be to take the renormalised mass
$m_R$ (which
however, requires a separate calculation). In numerical calculations
close to the critical point shown in Fig.[1] we have chosen for simplicity
a small positive value.
We found that the lower lying physical mass spectrum is not very
sensitive to the value
of $m_{kin}$ (this is not the case for higher lying masses).

\bigskip

We diagonalised the Hamiltonian on two lattices:
$\Lambda/\Delta p=3$ and $\Lambda/\Delta p =4$.
This would correspond to symmetric lattices ($-\Lambda$ and $+\Lambda$)
of size $7^3$ and
$9^3$ nodes, respectively. This results in
a very small Hilbert space of only 6 and 21 states, respectively.
In order to compare our results to those of L{\"u}scher and Weisz\cite{Luscher}
we express the bare parameters $m_0$
and $g_0$ in terms of the parameters $\lambda$ and $\kappa$:
$m_0^2=(1-2\lambda)/\kappa-8$ and
$g_0= 6{\lambda \over \kappa^2}$.
Fig.[1] displays the renormalised mass $m_R$ versus
$\kappa$. One observes that our results computed on very small lattices
are quite close to the results of L{\"u}scher and Weisz\cite{Luscher}.
Masses $M$ computed on the lattice must obey $a< 1/M < L$ where $L$
is the length of the lattice and $a$ denotes the lattice spacing of a
space-time lattice $\Lambda = {\pi \over a}$.
It can be shown from perturbation theory\cite{Brezin,Luscher} that
the physical masses close to the critical point
obey the following scaling law
 $M \sim C \tau^{1/2}|ln \tau|^{-1/6}$, where $\tau:=1-\kappa/\kappa_{crit}$
and $C$ is a constant. Since the results of Ref.\cite{Luscher} are based on the
solution of the renormalisation group equations, this scaling law
fits their results.
One should note, however, that two different regularisations
(this work and that of Ref.\cite{Luscher}) in general correspond to two
different critical lines. In Tab. [1] we have displayed our
results for the critical points $\kappa_{crit}$ as a function
of $\lambda$ and compared our results with those of Ref.\cite{Luscher}.
Again, our results are very close to those of L\"uscher and Weisz.
These results cover a domain of the bare parameter space extending
quite far away from the Gaussian fixed point $\kappa=1/8$ and $\lambda=0$.

\bigskip

Another way to test continuum physics is to look at the mass
ratios $M_n/M_1$ from the spectrum on the lattice and check if they
become independent of the cutoff $\Lambda$ or else independent of the
coupling constant $g_0(\Lambda)$ (i.e., they scale).
Those mass ratios $M_n/M_1$ are shown in Fig.[2]. As can be seen,
for a number of states $M_n/M_1\rightarrow const$
in a wide range of $\kappa$-values, i.e., they scale.
However, for some states $M_n/M_1$ diverges, i.e., there is no scaling.
The physical reason behind this is the following: The $\phi^4 _{3+1}$ model
describes a gas of partons repelling each other\cite{Luscher}.
The spectrum of Fig.[2] shows states dominated by the 1-,2-,3-,4- particle
Fock space sectors plus a spectrum of excited (scattering) states.
The picture of repulsive two-particle-exchange force is confirmed
by observation that the mass of the lowest-lying $n$-body state
is larger than $n$-times the mass of the one-body state.
The states which scale are just those lowest-lying $n$-body states.
The higher-lying part of the spectrum consits of states with
more nodes in the wave-function than lattice points, having also
a wider range and contributions from higher Fock-state sectors.
Because in the calculation corresponding to Fig.[2], the parameters
$\Delta p$, $\Lambda$ and the parton number cutoff are all kept fixed,
we cannot properly describe those higher-lying states.
Consequently,
they do not show scaling. When we go to bigger lattices
($\Delta p\rightarrow 0$) then we observe (not displayed here)
more states which show scaling.
\\
\\
{\Large \bf Distribution functions}
\\
The distribution function $f(x_B,Q)$ of finding some parton with
momentum fraction $x_B$ inside the hadron is determined by the
parton momentum distribution function $\tilde f(\vec p, \vec P)$ for
finding a parton with momentum $\vec p$ inside the hadron with momentum
$\vec P$. Since $Q$ is a dimensionful quantity, its scale is set by
the lattice spacing $a$, i.e., $Q \sim  \displaystyle{1\over a(m_0,g_0)}$
and thus depends on
the bare parameters if one keeps the
renormalised mass fixed. The continuum limit $a\rightarrow 0$ corresponds
to the the limit towards arbitrarily high resolution ability. If one
keeps the renormalised mass and the renormalised coupling constant fixed,
then $Q$ is a function of the bare coupling constant $g_0$ and vice
versa -- invertibility of $Q(g_0)$ assumed. Hence $f(x_B,Q)$ is related via
the function $Q(g_0)$ to
the distribution function $\bar f(x_B,g_0)$ which only depends on dimensionless
parameters. Consequently, a calculation
of the distribution function along a renormalisation group trajectory
can be used to compute the $Q$-dependence
 of the quark structure functions in $QCD$.

\bigskip

While $QCD$ possesses bound states of quarks and gluons, the existence
of corresponding bound states in the scalar $\phi ^4 _{3+1}$ is not
evident. According to Ref.\cite{Luscher} they do not exist
in the symmetric phase and there is little chance to find them
in the broken phase, either. This is confirmed by our numerical findings.
In order to compute distribution functions of a bound state of partons
in the scalar model we have taken recurrence to the $\phi^3$ model.
We calculate the distribution function of the
$\phi^3$ theory, because the $\phi^3$-interaction describes forces which are
attractive one-particle exchange forces\cite{Luscher}. This allows formation of
bound states as in $QCD$.
However, this theory is known to suffer from an unstable vacuum since
it is unbounded from below.
The unstable vacuum of the $\phi^3$
theory prevents to calculate meaningful ground state masses which
are needed to specify renormalisation group trajectories and hence
the exact relation between the resolution $Q$ and the bare coupling constant
$g_0$.
While in $QCD$ one computes $\bar f(x_B,g_0$ and $g_0(Q)$ to obtain
$f(x_B)$, here we can only compute the distribution function $f(x_B,g_0(Q)$.
We have computed the distribution function in 1-,2- and 3 space dimensions.
For a given parton number cutoff, these curves look very much alike.
In order to analyse the behaviour at small $x_B$  we have chosen to present our
result corresponding to a calculation in one space dimension (Fig.[3]).
When increasing the coupling $g_0$ we see that the distibution function
develops a peak at momentum fraction $x_B=0$. This is so, because increasing
the coupling means that more partons are produced which share the total
momentum fraction. The behaviour of
the distribution function seen here is typical
for $QCD$, where $g_0(Q)$ increases
with the resolution $Q$.
It is seen in
$DIS$ experiments and described by the Altarelli-Parisi equations.
If we had applied a parton number cutoff independent of $\Lambda$, the
small $x_B$ behaviour of Fig.[3]
which is a typical many-body effect\cite{DiplomArbeit},
would not have been seen.
This is so because a system of $n$ identical observable particles
must have an expectation value of $x_B$ around $1/n$ for symmetry reasons.

In conclusion, we have devised a Hamiltonian method able to compute
physical observables in Minkowsky space. We have applied it to
the scalar model and obtained the correct scaling behaviour of
the mass spectrum at the critical point. Moreover, we have
computed distribution functions showing a peak at small $x_B$ as
described by the Altarelli-Parisi equations in $QCD$.
Work is in progress to compute structure functions for full $QCD$.

\acknowledgements

 One of the authors (N.Scheu) wants to express his appreciation for
having been granted the AUFE fellowship from the DAAD (Deutscher Akademischer
Austauschdienst)
which has made this Ph.D. project possible.


\begin{figure}

{\bf Fig.1}
The ground state mass $m_R$ in lattice units ($a\equiv 1$) versus $\kappa$
for $\lambda=0.00345739 $ ($\bar \lambda= 0.01$ in Ref.\cite{Luscher}).
The dots correspond to results of Ref.\cite{Luscher}. Our results
correspond to $\Lambda/\Delta p = 3$ (dashed line) and $\Lambda/\Delta p =4$
(solid line).
\end{figure}

\begin{figure}

{\bf Fig.2}
The lowest lying mass spectrum versus $\kappa$. The ground state mass
is set to one. $\lambda$ as in Fig.[1].
\end{figure}

\begin{figure}

{\bf Fig.3}
The distribution function $\bar f(x_B,g_0(Q))$ of $\phi^3 _{1+1}$ versus the
momentum
fraction $x_B$ and the coupling constant $g_0(Q)$. The bare mass $m_0$ has been
to be $m_0=3 \Delta k$. $\Lambda/\Delta p = 11$.

\end{figure}

\section{Table Caption}

 \begin{center}
 \begin{tabular}{|| c|c|c|c|c|c|c|c|c ||}\hline
\squeezetable
$\lambda$
 & 0.0005   & 0.001    & 0.005    & 0.01     & 0.05     & 0.1    \\
\hline
$\kappa^{LW} _{crit}$
& 0.125101 & 0.125202 & 0.125991 & 0.126968 & 0.132368 & 0.13601 \\
\hline
 $\alpha$
 & 0.99997  & 0.99993  & 0.99972  & 0.9993   & 1.0073   & 1.0275   \\
\hline
 \end{tabular}
 \end{center}

The critical points $\kappa_{crit}$ versus $\lambda$.
$\kappa_{crit} ^{LW}$ is taken from Ref.\cite{Luscher}.
$\alpha := \kappa^{KS} _{crit}/ \kappa^{LW} _{crit}$ denotes the
ratio between the results of this work and Ref.\cite{Luscher}.
In this work, $\kappa_{crit}$ has been determined by the condition
that the renormalised mass $m_R$ becomes imaginary. $\Lambda/\Delta p = 4$.

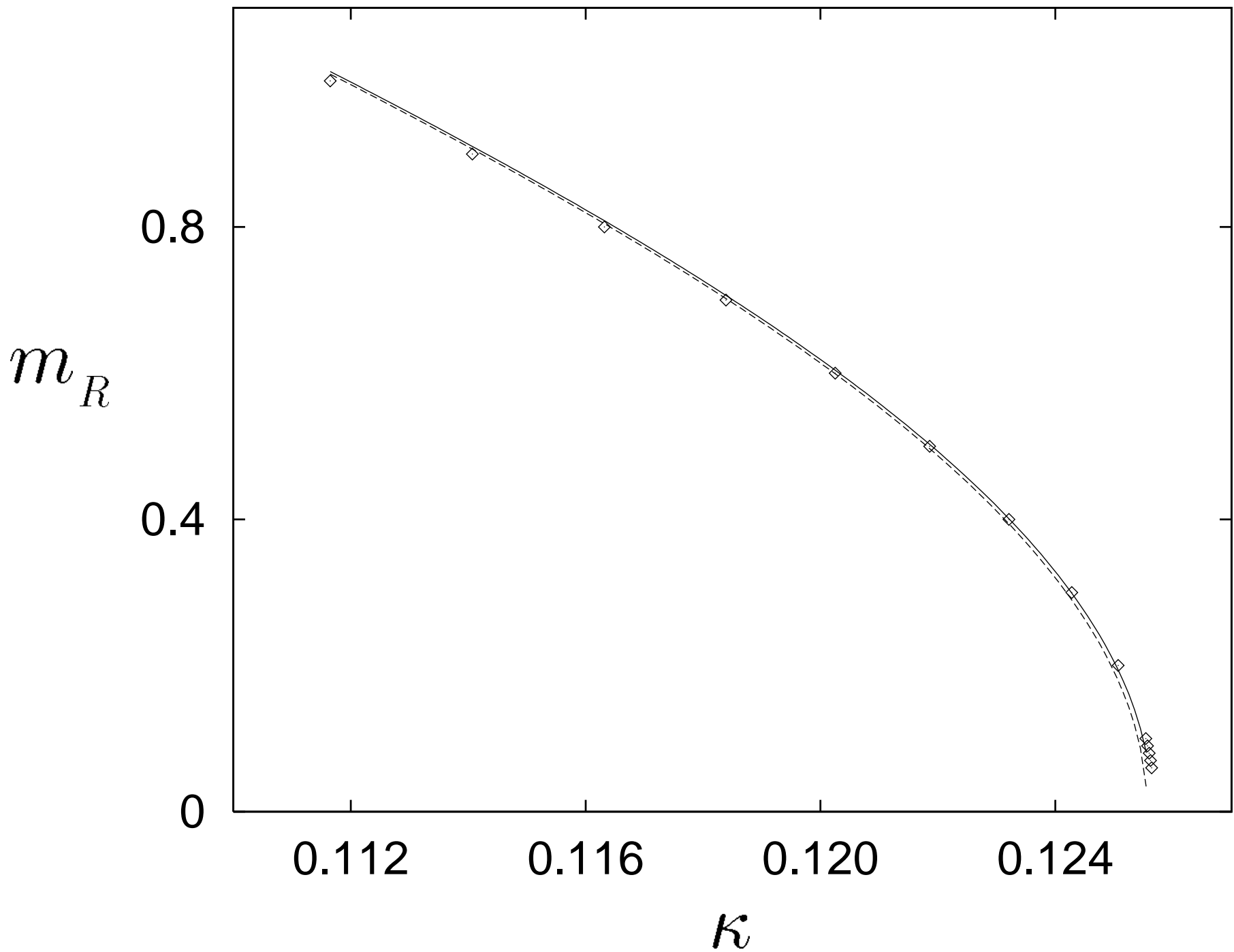
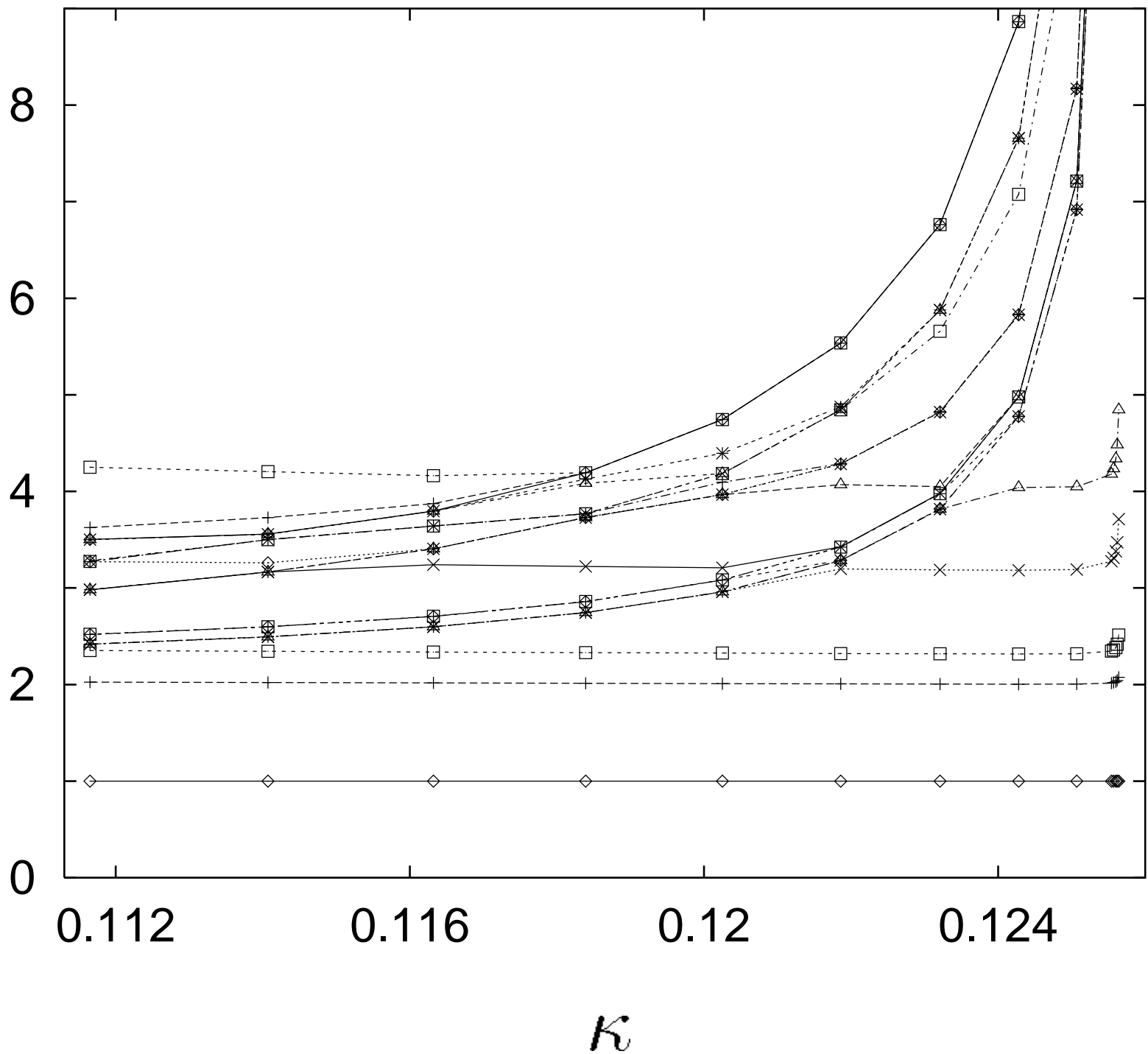
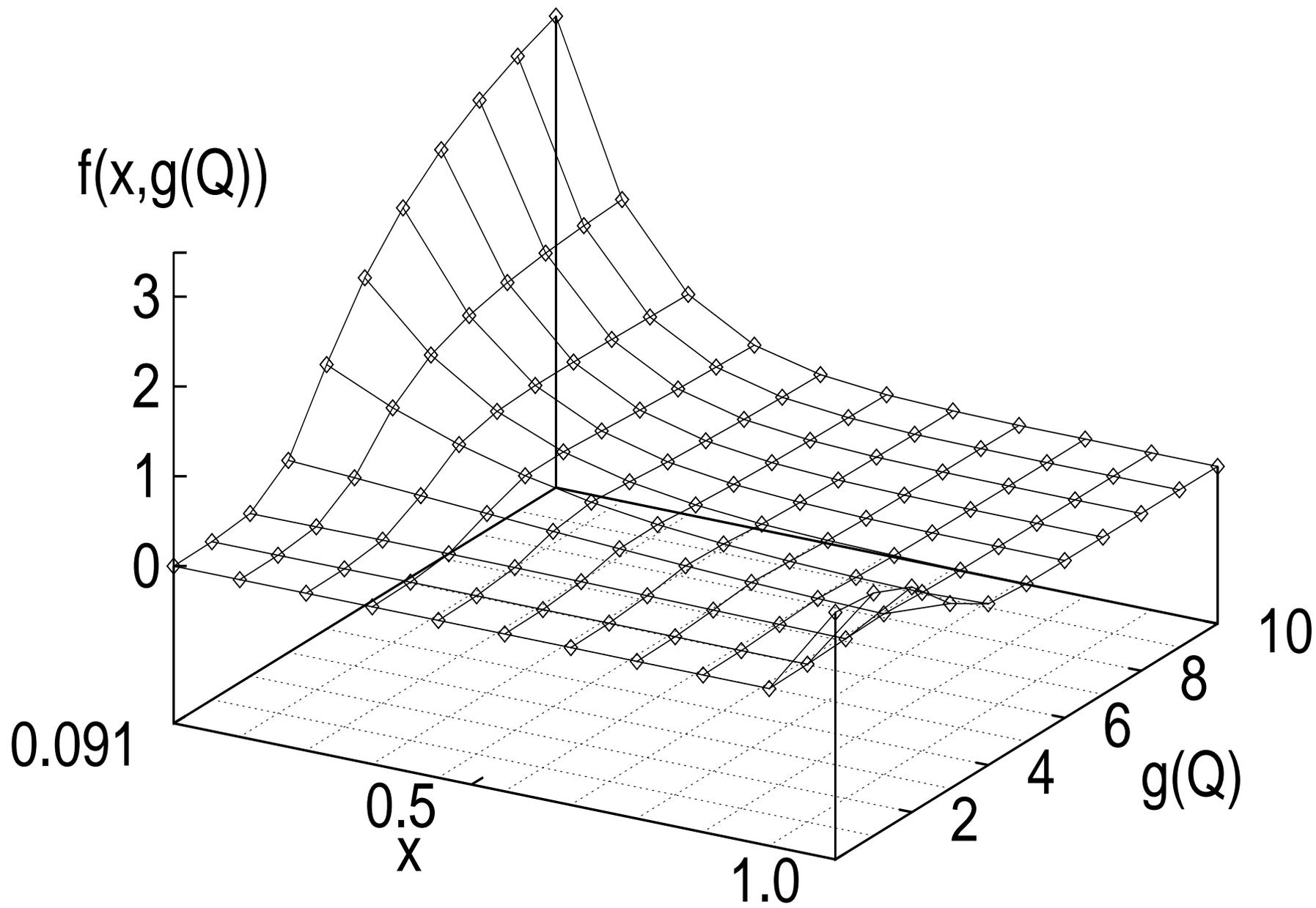
\end{document}